\documentclass[a4paper]{jpconf}
\usepackage{graphicx}
\usepackage{epstopdf}
\begin{document}
\title{Frame-Dragging from Charged Rotating Body}

\author{Anuj Kumar Dubey and A. K. Sen}

\address{Department of Physics, Assam University, Silchar-788011, Assam, India}

\ead{danuj67@gmail.com, asokesen@yahoo.com}

\begin{abstract}
   In the present paper, we have considered the three parameters: mass, charge and rotation to discuss their combined effect on frame dragging for a charged rotating body. If we consider the ray of light which is emitted radially outward from a rotating body then the frame dragging shows a periodic nature with respect to coordinate $\phi$ (azimuthal angle). It has been found that the value of frame dragging obtains a maximum at, $ \phi =\frac{\pi}{2}$ and a minimum  at $ \phi =\frac{3 \pi}{2}$.
\end{abstract}

\section{Introduction}

Considerations on rotation are important in general relativity and frame dragging is one of its examples. Frame-dragging is a general relativistic feature of all solutions to the Einstein field equations associated with the rotating masses. General relativity is dealing with the precession phenomena such as perihelion advancement precession, geodetic precession, Lense-Thirring precession \cite{lense1918} and Pugh-Schiff precession \cite{schiff1960}.

In 1960, L. I. Schiff \cite{schiff1960} showed that an ideal gyroscope in orbit around the Earth or other massive body would undergo two relativistic precessions with respect to a distant undisturbed inertial frame: (1) a geodetic precession in the orbit plane due to the local curvature of space time; (2) a frame-dragging precession due to the Earth's rotation.

After Pugh-Schiff precession, a proposal came for Gravity Probe-B (GP-B) in 1959 and finally launched in 2004, as an experimental confirmation of frame-dragging. The objective of Gravity Probe-B experiment was to design, develop, conduct, and analyze the data of a flight experiment to test the predictions of Albert Einstein's general relativity.

In GP-B experiment, Schiff's formula has been used for the combined gyroscope precession: $ \Omega =\frac{3G M}{2 c^{2}R^{3}}(\textbf{R}\times \textbf{V})+ \frac{ GI}{c^{2}R^{3}}  [\frac{3R}{R^{2}}(\omega.\textbf{R})-\omega]$. Here the first and second terms are the geodetic, and the frame-dragging effect. G and c are the gravitational constant and the velocity of light respectively.  M, I and $\omega$ are the mass, moment of inertia, and angular velocity of the earth respectively.  R and v are the instantaneous distance and velocity of the gyroscope respectively.

Everitt et al. in 2008 \cite{everitt2008} had confirmed that the geodetic effect (caused by the earth's mass; the amount by which the earth warps the local space-time in which it resides) and the frame-dragging effect (caused by the rotation of the earth; the amount by which the rotating earth drags its local space-time around with it). In the 642 km orbit of GP-B, the predicted geodetic and frame-dragging precessions are 6606 marcs/yr and 39 marcs/yr respectively \cite{everitt2008}.

Collas and Klein in 2004 \cite{collas2004} had showed that a prototype example of frame-dragging arises in the Kerr metric. A test Particle with zero angular momentum released from a non-rotating frame, far from the source of the Kerr metric, accumulates non-zero angular velocity in the same angular direction as the source of the metric, as the test particle plunges toward the origin. This dragging of inertial frames or frame-dragging, is due to the influence of gravity alone, and has no counterpart in Newtonian physics. Frame-dragging is a general relativistic feature, not only of the exterior Kerr solution, but of all solutions to the Einstein field equations associated with rotating sources. They also showed that surprising frame-dragging anomalies can occur in certain situations. They had considered the examples of axially symmetric solutions to the field equations in which zero angular momentum test particles, with respect to non-rotating coordinate systems, acquire angular velocities in the opposite direction of rotation from the sources of the metrics. They refer to this phenomenon as negative frame-dragging.

Mei and Jiu in 2006 \cite{mei2006} had showed that the angular velocity of frame-dragging is negative in the situation where non zero components of the metric $g_{\phi\phi}$ is positive. Under this situation, the zero angular momentum test particle may acquire an angular velocity in the opposite direction of rotation from the source of the metric. Such negative frame-dragging is also called frame-dragging anomaly in general relativity. Its occurrence is dependent on the coefficient of the metric in the $\phi$ coordinate. They also showed that negative frame-dragging occurs in non-rotating reference frames for the metrics of all solutions of the Kerr family which are associated with rotating sources.

Epstein in 2008 \cite{epstein2008} had discussed that a Hamiltonian approach makes the phenomenon of frame-dragging from the appearance of the drag velocity in the Hamiltonian of a test particle in an arbitrary metric. A distinction is made between manifest frame-dragging such as that in the Kerr metric, and the hidden frame-dragging that can be made manifest by a coordinate transformation such as that applied to the Robertson-Walker metric. They also showed that a zone of repulsive gravity is found in the extreme Kerr metric. They also explained the frame-dragging in special relativity as a manifestation of the equivalence principle in accelerated frames.

 Dubey and Sen in 2015 \cite{dubey2014} had obtained the expression of angular velocity of frame-dragging in Kerr Geometry by considering the ray of light emitted radially outward from the surface of a rotating compact object. They obtained the general expression of frame-dragging as a function of coordinates, $\phi$ (azimuthal) and $\theta$ (latitude). Dubey and Sen in 2015 \cite{dubey2015} had also obtained the general expression of angular velocity of frame-dragging in Kerr-Newman Geometry by considering a charged rotating body having electric charge (Q) and magnetic charge (P).
\section{Geometry of charged rotating body}

As a result of work carried out by R. H. Price, B. Carter, W. Israel, D. C. Robinson, and S. W. Hawking, black holes, from the point of view of the outside observer, can possess only three distinguishing characteristics: mass, charge and angular momentum.

When rotation is taken into consideration spherical symmetry is lost and off-diagonal terms appear in the metric and the most useful form of the solution of Kerr family (Kerr metric and Kerr-Newman metric) is given in terms of t, r, $\theta$ and $\phi$, where t, and r are Boyer-Lindquist coordinates running from - $\infty$ to + $\infty$, $\theta$  and $\phi$, are ordinary spherical coordinates in which  $\phi$ is periodic with period of 2$\pi $ and $\theta$ runs from 0 to $\pi$. Covariant form of metric tensor for Kerr family (Kerr 1963 \cite{kerr1963}, Newman et al. 1965 \cite{newman1965}) in terms of Boyer-Lindquist coordinates with signature (+,-,-,-) is expressed as:
\begin{equation}
ds^{2} = g_{tt}c^{2}dt^{2}+ g_{rr} dr^{2} + g_{\theta\theta} d\theta^{2} +g_{\phi\phi} d\phi^{2} + 2 g_{t \phi} c dt d\phi
\end{equation}
where $g_{ij}$'s are non-zero components of Kerr family. \\
As detailed in our previous work Dubey and Sen 2015 \cite{dubey2015}, if we consider the three parameters: mass (M), rotation parameter (a) (defined as angular momentum per unit mass setting speed of light unity), and charge (electric (Q) and / or magnetic (P)). Then it is easy to include charge in the non-zero components of $g_{ij}$ of Kerr metric, simply by replacing $(r_{g} r)$ with $(r_{g} r-Q^{2}-P^{2})$. Here r and $r_g=(2GM/c^2)$ are the radial coordinate and Schwarzschild radius respectively.

 Non-zero components of $g_{ij}$ of Kerr-Newman metric are given as follows (page 261-262 of Carroll 2004 \cite{carroll2004}; Dubey and Sen 2015 \cite{dubey2015}):
\begin{equation}
 g_{tt} = (1-\frac{r_{g} r-Q^{2}-P^{2}}{\rho^{2}})
\end{equation}
\begin{equation}
g_{rr}=-\frac{\rho^{2}}{\Delta}
\end{equation}
 \begin{equation}
 g_{\theta\theta}=-{\rho^{2}}
 \end{equation}
 \begin{equation}
g_{\phi\phi}= -[r^{2}+a^{2}+\frac{(r_{g}r-Q^{2}-P^{2}) a^{2}sin^{2}\theta}{\rho^{2}}]sin^{2}\theta
 \end{equation}
\begin{equation}
g_{t\phi}=\frac{a sin^{2}\theta (r_{g}r-Q^{2}-P^{2})}{\rho^{2}}
\end{equation}
with
\begin{equation}
\rho^{2} = r^{2}+ a^{2}cos^{2}\theta
\end{equation}
and
\begin{equation}
 \Delta =r^{2}+ a^{2}-r_{g}r +Q^{2}+P^{2}
\end{equation}
If we replace $(r_{g} r-Q^{2}-P^{2})$ by $(r_{g} r-Q^{2})$ and further if we put rotation parameter of the source (a) equal to zero, then it reduces to Reissner-Nordstrom metric. Also  if we replace $(r_{g} r-Q^{2}-P^{2})$ by $(r_{g} r)$ then the Kerr-Newman metric reduces to Kerr metric and further if we put rotation parameter of the source (a) equal to zero then it reduces to Schwarzschild metric.
\section{Frame-dragging from charged rotating body}

As detailed in our previous work Dubey and Sen 2015 \cite{dubey2015}, the Lagrangian $\pounds$ of the test particle can be expressed as:
\begin{equation}
\pounds  = \frac{g_{ij}\dot{x}^{i}\dot{x}^{j}}{2}
\end{equation}
From the Lagrangian of the test particle, we can obtain the momentum of the test particle as:
\begin{equation}
P_{i} = \frac{\partial\pounds}{\partial\dot{x}^{i}}= g_{ij} \dot{x}^{j}
\end{equation}
Thus corresponding momentum in the coordinates of t, r, $\theta$, and $\phi$ are given by
\begin{equation}
P_{t} \equiv - E= c^{2} g_{tt} \dot{t} + c g_{t\phi} \dot{\phi}
\end{equation}
\begin{equation}
P_{r} = g_{rr} \dot{r}
\end{equation}
\begin{equation}
P_{\theta} = g_{\theta\theta} \dot{\theta}
\end{equation}
\begin{equation}
P_{\phi} \equiv L= c g_{t\phi} \dot{t} + g_{\phi\phi} \dot{\phi}
\end{equation}
Since the space time of the Kerr family is stationary and axially symmetric, the momenta $P_{t}$ and $P_{\phi}$  are conserved along the geodesics. So we obtain two constants of motion: one is corresponding to the conservation of energy (E) and the other is the angular momentum (L) about the symmetry axis.\\
Using equations (11) and (14) we can write:
\begin{equation}
c\dot{t} =\frac{g_{\phi\phi} E + c g_{t\phi} L }{c g^{2}_{t\phi}- c g_{tt} g_{\phi\phi}}
\end{equation}
and
\begin{equation}
\dot{\phi} = -\frac{g_{t\phi} E + c g_{tt} L }{c g^{2}_{t\phi}- c g_{tt} g_{\phi\phi}}
\end{equation}
Using equations (15) and (16), we can obtain the angular velocity of frame-dragging ($\frac{d\phi}{cdt}$) as (similar expression was given by Collas and Klein 2004 \cite{collas2004}; Mei and Jiu 2006 \cite{mei2006}; Dubey and Sen 2015 \cite{dubey2014}):
\begin{equation}
\frac{\dot{\phi}}{c\dot{t}}=\frac{u^{\phi}}{u^{t}}= \frac{d\phi}{cdt} =- \frac{g_{t\phi} E + c g_{tt} L }{g_{\phi\phi} E + c g_{t\phi} L}
\end{equation}
Considering L=0, Collas and Klein 2004 \cite{collas2004}; Mei and Jiu 2006 \cite{mei2006}, had obtained angular velocity of frame-dragging of a zero angular momentum test particle as:
\begin{equation}
\frac{d\phi}{cdt}=- \frac{g_{t\phi} E  }{g_{\phi\phi} E}
\end{equation}
Considering the ray of light emitted radially outward from the surface of a compact object (from a rotating body with radius \lq R\rq), the general expression of angular velocity of frame-dragging ($\frac{d\phi}{cdt}$) in Kerr Geometry was given as (Eqn. (46) of  Dubey and Sen 2015 \cite{dubey2014}):
 \begin{equation}
\frac{d\phi}{cdt}( \phi, \theta)=\frac{\frac{R sin\phi sin\theta}{sin^{2}\theta} (1-\frac{r_{g}r}{\rho^{2}})+\frac{r_{g}ra}{\rho^{2}}}{-\frac{r_{g}r a}{\rho^{2}} (R sin \phi sin\theta) + (r^{2}+a^{2}+\frac{r_{g}r a^{2}}{ \rho^{2}}sin^{2}\theta)}
\end{equation}
Considering a rotating body having electric charge (Q) and magnetic charge (P), the general expression for angular velocity of frame-dragging ($\frac{d\phi}{cdt}(\phi, \theta)_{KN}$) for a charged rotating body in Kerr-Newman Geometry was given as (Eqn. (29) of  Dubey and Sen 2015 \cite{dubey2015}):
\begin{equation}
\frac{d\phi}{cdt}(\phi,\theta)_{KN}=\frac{\frac{R sin\phi sin\theta}{sin^{2}\theta} (1-\frac{(r_{g}r-Q^{2}-P^{2})}{r^{2}+ a^{2}cos^{2}\theta})+\frac{(r_{g}r-Q^{2}-P^{2})a}{r^{2}+ a^{2}cos^{2}\theta}}{-\frac{(r_{g}r-Q^{2}-P^{2}) a}{r^{2}+ a^{2}cos^{2}\theta} (R sin \phi sin\theta) + (r^{2}+a^{2}+\frac{(r_{g}r-Q^{2}-P^{2}) a^{2}}{ r^{2}+ a^{2}cos^{2}\theta}sin^{2}\theta)}
\end{equation}
In the above expression of frame-dragging given by equation (20):
\begin{itemize}
  \item If we substitute P=0 and a=0 then we can obtain the coresponding expression of frame-dragging in  Reissner-Nordstrom Geometry.
  \item If we substitute Q=P=0, then we can obtain the coresponding expression of frame-dragging in Kerr Geometry.
 \item If we substitute Q=P=0 and a=0, then we can obtain the coresponding expression of frame-dragging in Schwarzschild Geometry, which can be considered as Geodetic effect.
\end{itemize}
If we substitute $\theta =\frac{\pi}{2}$ (equator) in equation (20). We can obtain the expression of frame-dragging $(\frac{d\phi}{cdt})$ for the equatorial plane in the Kerr-Newman Geometry as:
\begin{equation}
\frac{d\phi}{cdt}(\phi,\theta =\frac{\pi}{2}) = \frac { [1-(\frac{r_{g}}{r}-\frac{Q^{2}+P^{2}}{r^{2}})]( R sin\phi)+a(\frac{r_{g}}{r}-\frac{Q^{2}+P^{2}}{r^{2}})}{-a(\frac{r_{g}}{r}-\frac{Q^{2}+P^{2}}{r^{2}}) (R sin\phi) + (r^{2}+a^{2}+a^{2}(\frac{r_{g}}{r}-\frac{Q^{2}+P^{2}}{r^{2}}))}
\end{equation}

If we substitute $\theta =\frac{\pi}{2}$ (equator) and  charge Q=P=0 in equation (20). We can obtain the expression of frame-dragging $(\frac{d\phi}{cdt})$ for the equatorial plane in the Kerr Geometry as:
\begin{equation}
\frac{d\phi}{cdt}(\phi,\theta =\frac{\pi}{2}) = \frac { (1-\frac{r_{g}}{r})( R sin\phi)+\frac{r_{g}a}{r} }{-\frac{r_{g} a}{r} ( R sin\phi) + (r^{2}+a^{2}+\frac{r_{g} a^{2}}{r})}
\end{equation}
\begin{figure}
\includegraphics[width=40pc, height=40pc,angle=270]{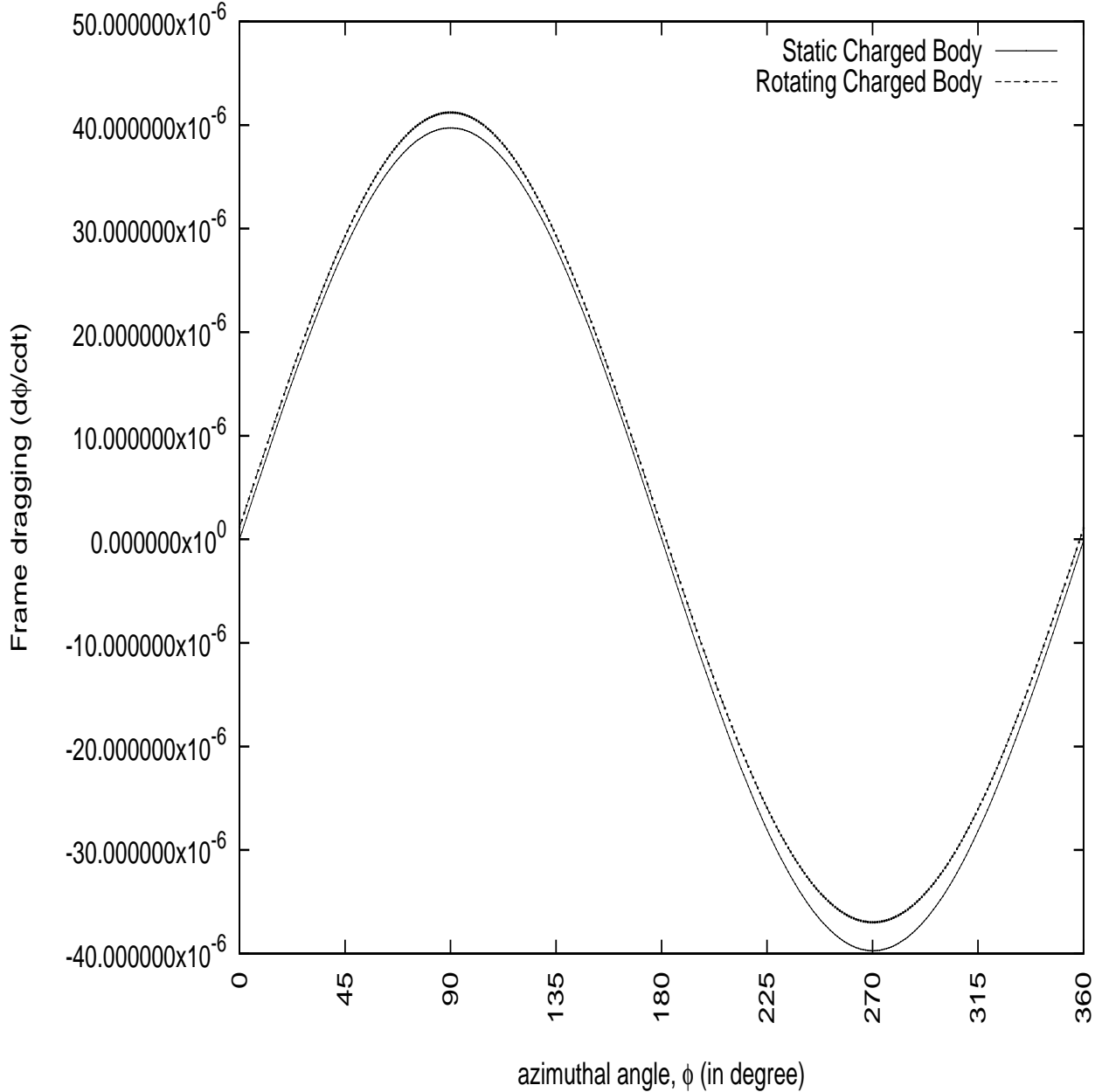}
 \caption{\label{fig1} Shows the variation of frame-dragging $\frac{d\phi}{cdt}(\phi,\theta =\pi/2)$ versus $\phi$ for pulsar PSR J1748-2446 ad, from $\phi$ = 0 to $2\pi$,  at fixed $\theta$ = $\pi$ and at fixed value of charge.}
\end{figure}
For doing numerical calculations, we have considered a rotating star such as pulsar PSR J 1748-2446ad (Hessels et al. 2006 \cite{hessels2006}), having mass (M= 1.35 Solar Mass), physical radius (R= 20.10 km), Schwarzschild radius ($r_{g}$= 4.05 km), angular velocity of rotation ($\omega= 4.50\times10^{3}$ rad/s),  rotation parameter (a = 2.42 km) and  a fixed value of charge.

Using equation (21), we made a plot (Figure-1) which shows the variation of frame-dragging $(\frac{d\phi}{cdt}(\phi, \theta=\pi/2)$ versus azimuthal angle $(\phi)$ for PSR J 1748-2446ad from $\phi$= 0 to 2$\pi$, at fixed $\theta(=\pi/2)$ and fixed value of charge. Here in the plot, solid line represents the frame-dragging for charged static body while dashed line for charged rotating body of same mass.

From Figure-1, it is clearly seen that the frame-dragging from a rotating star (as pulsars) shows a periodic nature with respect to coordinate $\phi$ (azimuthal angle). The value of frame-dragging is maximum at $\phi= \pi/2 $ and minimum at $\phi=3\pi/2$.  If we compare the frame-dragging for static body  and rotating body of same mass and charge, then we notice that the effect of rotation is to increase the value of frame-dragging. It may be also noted that, the effect of charge on frame-dragging is not very significant.\\

Using the concept of frame-dragging, the general expression of gravitational redshift in Kerr and Kerr-Newman geometry have been discussed in details in our previous works Dubey and Sen 2015 \cite{dubey2014}; Dubey and Sen 2015 \cite{dubey2015}.

\section{Conclusions}

We can conclude from the present work that,
\begin{enumerate}
\item Frame-dragging is a function of the azimuthal angle ($\phi$) and latitude ($\pi/2 - \theta$).
\item  Frame-dragging is mainly affected by the rotation of the central body.
\item  Frame-dragging is also affected by the electric and magnetic charge present on the rotating body.
\item   The effect of charge on frame-dragging is not very significant, in comparison to the rotation effect on frame-dragging.
\item If we consider the ray of light which is emitted radially outward from a rotating star then the frame-dragging shows a periodic nature with respect to coordinate $\phi$ (azimuthal angle).
\item  The value of frame-dragging obtains a maximum at, $ \phi =\frac{\pi}{2}$ and a minimum  at $ \phi =\frac{3 \pi}{2}$.
\end{enumerate}
\section{Acknowledgement}

We wish to thank Dr. Atri Deshmukhya,  Department of Physics, Assam University, Silchar, India for useful suggestions and inspiring discussions. 

\end{document}